\documentclass{edp-conf}
\usepackage{graphicx}
\usepackage{natbib}
\input{psfig}                              %

\def\Msol{\mbox{ }M_{\odot}}
\def\cmmoinsdeux{\mbox{ cm}^{-2}}
\def\grs{\mbox{GRS 1915+105}}
\def\gx{\mbox{GX 339-4}}
\def\xtejodh{\mbox{XTE J1118+480}}
\def\deg{^{\circ}}

\begin{document}

\TitreGlobal{SF2A 2001}

\title{Multiwavelength observations revealing 
the outbursts of the two soft X-ray transients
XTE J1859+226 and XTE J1118+480} 
\runningtitle{Multiwavelength observations of XTE J1859+226 and XTE J1118+480}
\author{S.~Chaty}\address{Department of Physics and Astronomy, 
The Open University, Milton Keynes, UK, s.chaty@open.ac.uk}
\author{C.A.~Haswell$^1$}
\author{R.I.~Hynes}\address{Department of Physics and Astronomy, 
University of Southampton, Southampton, UK}
\author{C.R.~Shrader}\address{Laboratory for High-Energy Astrophysics, 
NASA GSFC, Greenbelt, USA}
\author{W.~Cui}\address{Department of
Physics, Purdue University, West Lafayette, IN 47907, USA}%
\maketitle
\begin{abstract} 
We report multiwavelength observations of the two soft
X-ray transients (SXTs) XTE J1859+226 and XTE J1118+480, which we
observed with {\it HST}, {\it RXTE} and UKIRT.
The two sources exhibited very different behaviour. 
XTE J1859+226 showed a thermal-viscous disc instability outburst
modified by irradiation.
XTE J1118+480, which we also observed with {\it EUVE} since it is located 
at a very high galactic latitude and suffers from very
low extinction, is much more unusual.
It exhibits i) a low X-ray to optical flux ratio and ii) 
a strong non-thermal contribution throughout the spectrum, 
which is likely to be due to synchrotron emission.
We concentrate here on their evolution in the course of their outbursts.
 \end{abstract}
%
\section{Introduction}

SXTs, also called X-ray novae, 
are a class of low mass X-ray binaries (LMXBs), including
GRO J1655--40 and GRO J0422+32. 
More than 70\% of SXTs are thought to contain black holes \citep{charles:1998}.
The compact object accretes
matter through an accretion disc 
from a low-mass star via Roche lobe overflow. The history
of these sources is characterized by long periods of quiescence, 
typically lasting decades, and punctuated by very dramatic 
outbursts, visible at every wavelength. These
sources are usually discovered in X-rays or the optical, 
and often exhibit radio activity. 
Two such sources were discovered respectively in 1999 and 2000: XTE J1859+226
and XTE J1118+480.
Thanks to our pre-approved override programs on {\it RXTE}, {\it HST} and UKIRT 
we could get 
early multiwavelength observations of these systems,
and follow their evolution from outburst towards quiescence.

\section{XTE J1859+226} \label{1859}

The first source, XTE J1859+226, was discovered by ASM/{\it RXTE} on 1999
October 9 
at the galactic coordinates: ($l,b$) = ($54.05 \deg$, $+8.61 \deg$)
\citep{wood:1999}.
This source exhibited a fast rise ($\sim 5$ days) and 
exponential decay ($\sim 23$ days) typical of SXTs.
The optical counterpart reached 15th magnitude at its maximum
\citep{garnavich:1999} exhibiting a period of 9.15 +/-
0.05 hr \citep{garnavich:2000}, later shown to be the orbital
period by \citet{filippenko:2001}, who also determined the mass function 
of this object: $f(M) = 7.4 \pm 1.1 \Msol$ (the highest known), 
implying the compact object is a black hole.

The early spectral energy distribution (SED) which we observed 
(see Fig. 1) is well fitted
by a typical X-ray irradiated disc model ($T \propto R^{-3/7}$).
The model used was actually generated to fit the SED of GRO J0422+32
in outburst, which has an orbital period of 5.1 hr, and then
scaled to fit the new data with no other adjustment.
If the disc were heated by viscous processes instead of irradiation
we would expect intead to see $f_{\nu} \propto \nu^{1/3}$ 
(corresponding to $T\propto R^{-3/4}$).
Gratifyingly, this is seen in our last visit where the SED is
better fitted by a viscously heated accretion disc model 
with an edge temperature of $\sim 8000$ K,
suggesting evolution from an irradiation dominated to viscosity
dominated regime.

\section{XTE J1118+480} \label{1118}

The second source, XTE J1118+480, was discovered by {\it RXTE} on 2000 March 29 
\citep{remillard:2000} as a weak, slowly rising source, 
the post-analysis revealing
an outburst in January 2000, with a similar brightness.
The optical counterpart is a 13th magnitude star, coincident with a 18.8 mag
object in the DSS \citep{uemura:2000}.
This system was characterized by a very low X-ray to optical flux ratio
of 5 \citep{uemura:2000}; 
the typical value is 500 (see e.g. \citealt{tanaka:1996}).
A weak photometric modulation of 4.1 hr (0.17082 d) 
period was rapidly discovered \citep{cook:2000}, which was associated
with the orbital period, the shortest among the black hole
candidates. 
Flickering with an amplitude of $\sim 0.4$ mag,
and also a quasi-periodic oscillation (QPO) 
at 10 s, was observed in the optical, in the UV
\citep{haswell:2000c} and also in the X-rays with an evolving frequency
\citep{wood:2000}.
The large value of the mass function, $f(M) = 5.9 \pm 0.4 \Msol$,
implies the compact object is a black hole 
(\citealt{wagner:2000} and \citealt{mcclintock:2001a}).
The location of this object at a high galactic latitude
($l,b$) = ($157.62\deg$,$+62.32\deg$) is very unusual, 
and there is a very low absorption along the line of sight
of the source, with a column density estimated to
$N_{H} \sim 1.0-1.6 \times 10^{20} \cmmoinsdeux$ (see \citealt{hynes:2000}, 
Chaty et al. 2001, 
hereafter respectively H00 and C01).

Our unprecedented broadband 
coverage of the SED, shown in Fig. 2 
(see also H00 and C01),
suggests that the system was
exhibiting a low-state mini-outburst, with the inner radius
of the accretion disc at $\sim 200-300 R_s$ ($R_s$: Schwarzschild radius).
The SED shows a flat power-law spectrum from the radio to the NIR,
from the NIR to the UV, and in X-rays,
suggesting that there is another source of flux apart from thermal
disc emission (see also \citealt{markoff:2001}).
Furthermore, the SED did not evolve much during the
3 months of our coverage, a behaviour similar to that of 
jet sources such as $\grs$ and $\gx$ \citep{tanaka:1996}.
We detected flickering at NIR wavelengths, of bigger
amplitude ($\sim 0.8$ mag) than in the optical ($\sim 0.4$ mag).
All these facts, combined with the $\sim 10$ s QPO seen in the optical, 
UV and X-rays, suggest a strong {\bf non-thermal (likely synchrotron) emission} 
present from the radio to the X-rays (C01). \\

{\bf Acknowledgements:} 
S.C., C.A.H. and R.I.H. gratefully acknowledge
support from grant F/00-180/A from the Leverhulme Trust.

   \bibliographystyle{/usr/home/schaty/TeX/Bst/aa}

\newpage

\begin{figure}
\centerline{\psfig{file=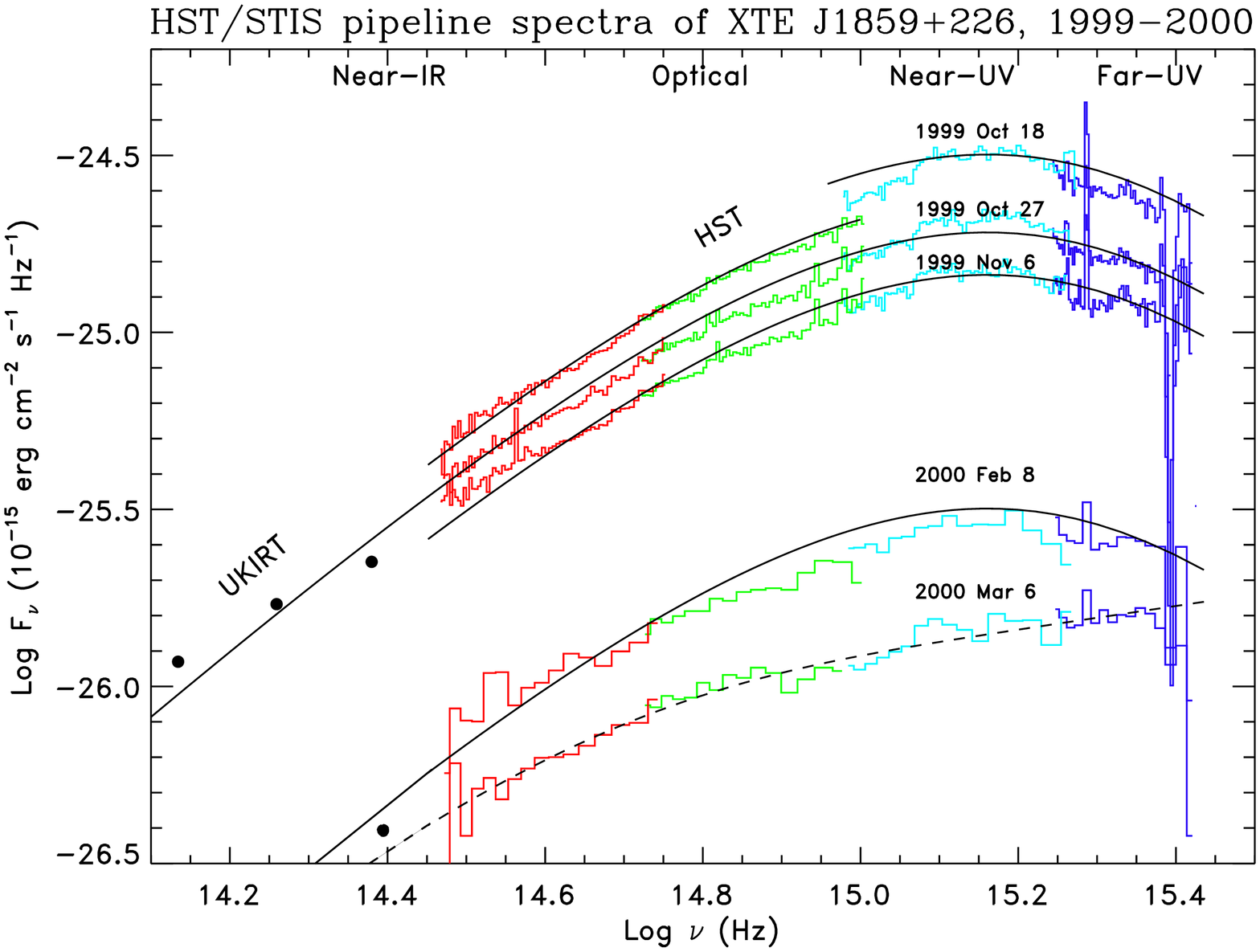,angle=+90.,width=11.cm}}
{\bf Figure 1 -- \small From irradiated to viscously-heated:}
{\small \it an irradiated spectrum ($T \propto R^{-3/7}$) fits the SED
during the outburst (top curves) while a viscously heated disc 
($T\propto R^{-3/4}$) matches the bottom, near quiescent SED.} 
\label{1859_ir}
\end{figure}

\begin{figure}
\centerline{\psfig{file=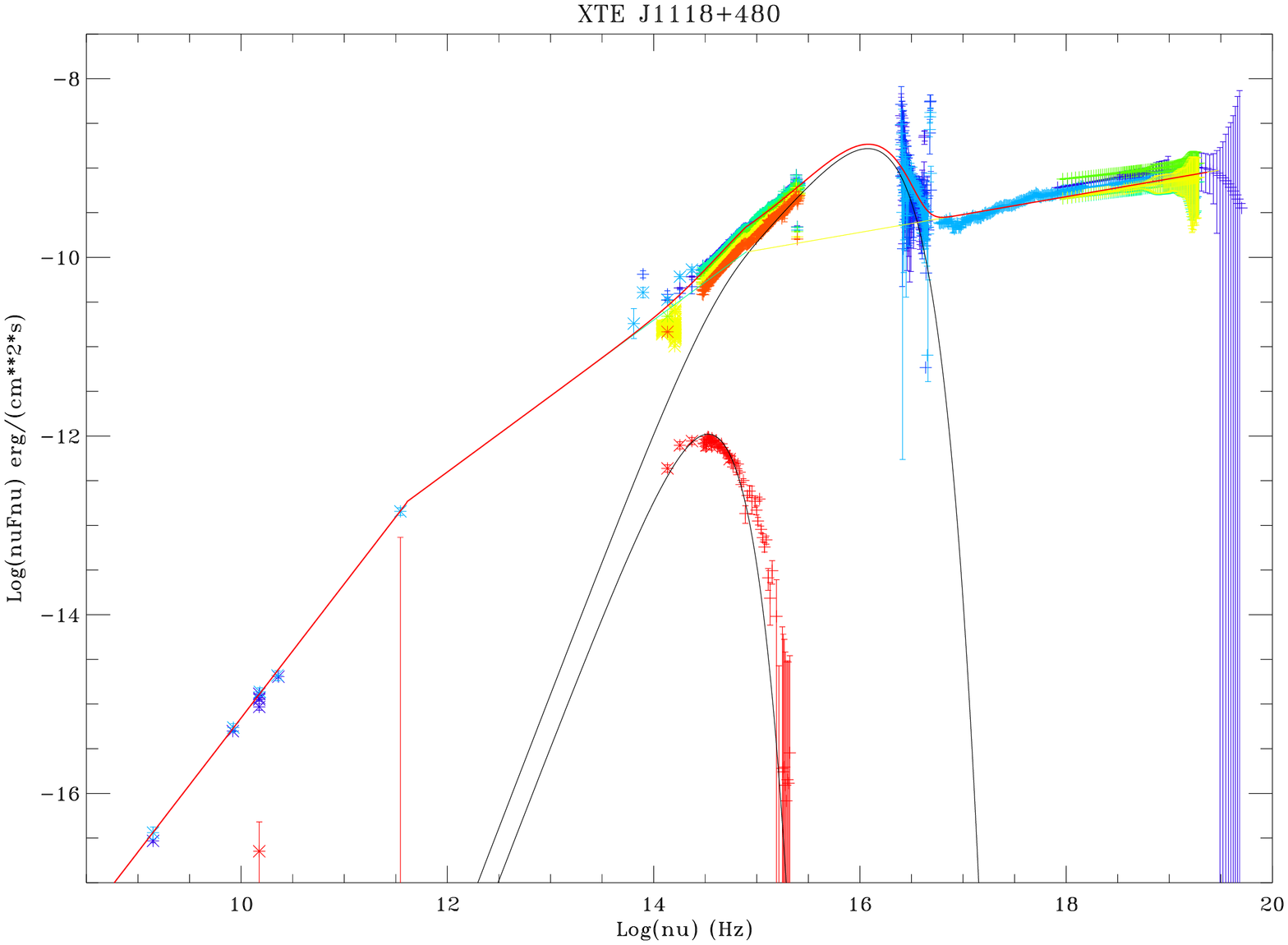,angle=90.,width=8.5cm}}
{\bf Figure 2 -- \small Spectral Energy Distribution:}
{\small \it Shown here are all the observations of $\xtejodh$ in an interval of 3 months.
The last HST and UKIRT observations of the source in near-quiescence
are reported at the bottom of the Figure, fitted by a black
body representing the companion star. 
The fluxes are corrected from interstellar absorption
with $N_H = 1.0 \times
10^{20} \cmmoinsdeux$. 
The model fitting approximatively all the observations
in a period of 3 months is the sum of: one steady-state
disc model with an outer disc at 8000K and inner disc
radius at $350 R_s$, 3 power laws of spectral indices respectively
0.5, -0.15 and -0.8 and the black-body representing the companion star
(see also C01 and H00).} \label{1118_sed}
\end{figure}

\end{document}